\documentclass[useAMS,usenatbib]{mn2e}
\usepackage{graphicx}
\usepackage{verbatim} 
\usepackage{color} 
\input epsf
\newcommand{\Vvec}{\mbox{\bf V}}

\newcommand{\vvec}{\mbox{\bf v}}

\newcommand{\xvec}{\mbox{\bf x}}
\newcommand{\yvec}{\mbox{\bf y}}

\newcommand{\zvec}{\mbox{\bf z}}

\newcommand{\Omegavec}{\mbox{\boldmath $\Omega$}}

\title[Non-axisymmetric vertical shear and convective instabilities]{Non-axisymmetric vertical shear and convective instabilities as a mechanism of angular momentum transport}
\author[Francesco Volponi]{Francesco Volponi\thanks{email: volponi@ppl.k.u-tokyo.ac.jp} 
\\ {Graduate School of Frontier Science, The University of Tokyo, Chiba 277-8561, Japan}}

\begin{document}

\date{}
\maketitle
\begin{abstract}
Discs with a rotation profile depending on radius and height are subject to an axisymmetric linear instability, the vertical shear instability. Here we show that non-axisymmetric perturbations, while eventually stabilized, can sustain huge exponential amplifications with growth rate close to the axisymmetric one. Transient growths are therefore to all effects {genuine} instabilities. {The ensuing} angular momentum transport is positive. These growths occur when the product of the radial times the vertical wavenumbers (both evolving with time) is positive for a positive local vertical shear, or negative for a negative local vertical shear. \\
We studied, as well, the interaction of these vertical shear induced growths with {a convective instability.  The asymptotic behaviour depends on the relative strength of the axisymmetric vertical shear ($s_{\rm vs}$) and convective ($s_{\rm c}$) growth rates.\\
For $s_{\rm vs}>s_{\rm c}$ we observed the same type of behaviour described above - large growths occur with asymptotic stabilization. When $s_{\rm c}>s_{\rm vs}$ the system is asymptotically unstable, with a growth rate which can be slightly enhanced with respect to $s_{\rm c}$.\\
The most interesting feature is the sign of the angular momentum transport. This is always positive in the phase in which the vertical shear driven transients growths occur, even in the case $s_{\rm c}>s_{\rm vs}$ .\\
Thermal diffusion has a stabilizing influence on the convective instability, specially for short wavelengths.}
\end{abstract}
\clearpage
\section{Introduction}
{The evolution of accretion discs  is determined by the outward transport of angular momentum, which induces infall of matter toward the orbital center. It is generally assumed that the accretion rate depends on an effective viscosity (Shakura \& Sunyaev 1973).  In  magnetized discs this mechanism is triggered by a linear axisymmetric destabilization, the magnetorotational instability, which occurs when a weak poloidal magnetic field is coupled to rotation (Balbus \& Hawley 1991).}\\
In hydrodynamic Keplerian discs, instead, the presence of an efficient outward transport of angular momentum is still an open question. In this context there is no linear axisymmetric mechanism of growth, since Kepler rotation is stable according to Rayleigh criterion.\\
A possibility is given by the bypass scenario (Ioannou \& Kakouris 2001, Chagelishvili et al. 2003), where arbitrarily  large amplifications of perturbations can occur due to the leading-trailing evolution of shearwise wavenumbers. These growths occur independently of the presence of a linear instability drive. This linear mechanism and non-linear interactions, providing a positive feedback loop, could trigger a subcritical transition into turbulence. However, Lesur \& Longaretti (2005) showed that, while subcritical transition to turbulence indeed occurs in non-stratified discs, values of the Shakura-Sunyaev $\alpha$ parameter (Shakura \& Sunyaev 1973) are much too low to give rise to an efficient  turbulent transport.

If radial stratification is taken into account radial entropy gradients can drive non-linear subcritical baroclinic destabilization leading to positive angular momentum transport  (Klahr \& Bodenheimer 2003, Lesur \& Papaloizou 2010) and formation of vortices.\\
However, discs are in general radially and vertically stratified and the angular velocity is a function of both radial and vertical coordinates (Kippenhahm \& Thomas 1982, Urpin 1984). Vertical velocity shear drives an axisymmetric instability (Goldreich \& Schubert 1967; Urpin 2003) which can result in positive transport of angular momentum in the non-linear regime (Arlt \& Urpin 2004). The linear destabilization mechanism was derived by means of a local dispersion relation in the full cylindrical geometry. Instability occurs when $k_r\gg k_z$. \\
In a recent study Nelson, Gressel \& Umurhan (2012) found that in non-axisymmetric non-linear simulations outward transport can occur with $\alpha\sim10^{-4}$.

Vertical convection is another possible path to turbulence. Ryu \& Goodman (1992) showed, by means of a linear analysis, that an unstable vertical stratification is not favourable for accretion; angular momentum transport occurs, but inward. Stone \& Balbus (1996) extended the analysis to non-linear regimes with analogous results. {Departures from this picture are however possible as shown by Lesur \& Ogilvie (2010), in the strongly non-linear regime of high Rayleigh number, and by Volponi (2010), for linear non-axisymmetric Boussinesq perturbations in discs with radial and vertical stratification.}

In the present study we will concentrate on the vertical shear and vertical convective instabilities. Precisely, first, we will extend the linear analysis of the vertical shear instability to non-axisymmetric perturbations. Results will be derived by solving the shearing sheet equations in the presence of vertical velocity shear (Knobloch \& Spruit 1986) and in the short wavelength regime. We will show that, while non-axisymmetric perturbations eventually decay, they experience transient exponential growths which are indistinguishable from a {genuine} instability.\\
Second, we will study the interaction between {the vertical shear and convective instabilities}. This is the central part of the present study. We will show that the presence of vertical shear changes the direction of the vertical convection induced angular momentum transport from inward to outward depending on the sign of the product of vertical and radial wavenumbers (i.e. positive for $A_z>0$ or negative for $A_z<0$, where $A_z$ is the local vertical shear). Thermal diffusion has a stabilizing influence on the convective instability, specially for short wavelengths.
\section{Equations and equilibrium}
In the shearing sheet approximation, the equations governing the dynamics of a 3-dimensional disc are
\begin{equation}
\partial_{t}{\cal D}+\nabla \cdot{\cal D \Vvec}=0,
\label{eq1.1}
\end{equation}
\begin{equation}
\partial_{t}{\Vvec}+\Vvec\cdot\nabla{\Vvec}=-\frac{\nabla{P}}{\cal D}-2{\Omegavec}\times{\Vvec}+2q{\Omega}^2 x \hat{{\xvec}}-{\Omega}^2 z \hat{{\zvec}},
\label{eq1.2}
\end{equation}
\begin{equation}
\partial_{t}{(\ln{S})}+\Vvec\cdot\nabla{(\ln{S})}=0,
\label{eq1.3}
\end{equation}
where $\cal D$ and $P$ are density and pressure, $\Vvec$ is the fluid velocity, $S=P{\cal D}^{-\gamma}$ is a measure of the fluid entropy, $\gamma$ is the adiabatic index, $\Omega$ is the local rotation frequency and $q$ is the shear parameter ($q=1.5$ for Keplerian rotation). The term $-2{\Omegavec}\times{\Vvec}$ is the Coriolis term, $2q{\Omega}^2 x \hat{{\xvec}}$ is the tidal expansion of the effective potential and $-{\Omega}^2 z \hat{{\zvec}}$ is the vertical gravitational acceleration. The equations are expressed in terms of the pseudo-Cartesian coordinates $x=r-r_0$, $y=r_0(\phi-{\phi}_0)$ and $z$  ($r_0$ and ${\phi}_0$ are reference radius and angle).\\
The above equations are more manageable than the full equations in cylindrical geometry due to the neglect of curvature effects. The simplification holds if the length scale of radial gradients $L\ll r$, where $r$ is the cylindrical radial coordinate (Knobloch \& Spruit 1986).\\
We consider fully stratified (i.e. radially and vertically) baroclinic discs. $P_{\rm e}(x,z)$ and ${\rho}_{\rm e}(x,z)$ are respectively the equilibrium pressure and density. Equilibrium in general requires the angular velocity to be $x$ and $z$ dependent. \\
The equilibrium velocity field is then given by
\begin{equation}
[-q\Omega x+\frac{\partial_{x}P_{\rm e}(x,0)}{2 \Omega {\rho}_{\rm e}(x,0)}+a_V z^2]\hat{\yvec},
\label{eq1.4}
\end{equation}
which is obtained from the expansion of the angular velocity profile about the midplane as given in Kley \& Lin (1992) (see as well Urpin 1984), in the limit $L/r\ll 1$. The dimensional coefficient $a_V$ can be derived from equation (A23) of Kley \& Lin (1992).\\
The $z$ dependence gives rise to vertical shear at the origin of the axisymmetric vertical shear instability. Notice that at the midplane the vertical shear goes to zero, so the instability drive is present away from it. This is the most interesting region for the present analysis. Away from the midplane we can approximate the last $z$-quadratic term, $a_V z^2$,  in~(\ref{eq1.4}) with a linear one, $\bar{A}_z z$. Therefore in the following our reference equilibrium flow will be
\begin{equation}
{\Vvec}_{\rm e}(x, z)=[-q\Omega x+\frac{\partial_{x}P_{\rm e}(x,0)}{2 \Omega {\rho}_{\rm e}(x,0)}+\bar{A}_z z]\hat{\yvec},
\label{eq1.6}
\end{equation}
bearing in mind that our conclusions will not hold at the midplane.
The second term on the RHS of the above equation will be dealt with as in  Johnson \& Gammie (2005) by considering the background flow as giving an effective shear rate
\begin{equation}
\tilde{q}(x){\Omega}=-\frac{d{V}_{\rm e}(x)}{dx},
\label{eq1.7}
\end{equation}
that varies with $x$.\\
Vertical equilibrium gives
\begin{equation}
\frac{\partial_{z}{P_{\rm e}}}{\rho_{\rm e}}=-\Omega^2 z \equiv -g_z.
\label{eq1.8}
\end{equation}
By defining $1/{L_P}_z={\partial_{z}P_{\rm e}}/{\gamma P_{\rm e}}$ we obtain from~(\ref{eq1.8})
\begin{equation}
g_z=-\frac{{c_s}^2}{{L_P}_z},
\label{eq1.10}
\end{equation}
where ${c_s}^2=\gamma P_{\rm e} /\rho_{\rm e}$.
\section{Linear perturbations}
We decompose the physical variables in equilibrium and perturbation parts
\begin{equation}
{\Vvec}={\Vvec}_{\rm e}+{\vvec}', \hspace{0.7cm} {\cal D}={\rho}_{\rm e}+{\rho}', \hspace{0.7cm} P= P_{\rm e} + P',
\label{eq1.11a}
\end{equation}
and consider the linearized equations.
Localized on the $x$ and $z$ dependent flow (see Johnson \& Gammie 2005) and on the $x$ and $z$  dependent density and pressure backgrounds, we consider short wavelength Eulerian perturbations of the type
\begin{equation}
{\delta}'(t,x,y,z)=\hat{{\delta}'}(t)e^{i\int \tilde{K}_x(t,x)dx+iK_yy+i\tilde{K}_z(t)z},
\label{eq1.11}
\end{equation}
where 
\begin{equation}
\tilde{K}_x(t,x)={K_x}+\tilde{q}(x)\Omega K_y t,  \hspace{0.2cm} \tilde{K}_z(t)={K_z}-\bar{A}_z K_y t.
\label{eq1.12}
\end{equation}
Notice that due to  the presence of vertical shear the vertical wavenumber as well evolves with time. The $x$ dependence in $\tilde{q}(x)$ is very weak and in the rest of the paper we will consider the effective shear parameter constant and  very close to its Keplerian value.\\
As previously stated to be consistent with the neglect of the curvature terms we consider a background with radial and vertical  length scales $L\ll r$ and $H\ll r$. The short wavelength perturbations are such that $\tilde{K}_x L \gg 1$ and $\tilde{K}_z H \gg 1$. \\
With equation~(\ref{eq1.11}) the evolution of linearized perturbations is given by 
\begin{equation}
\partial_{t}\frac{\hat{\rho'}}{\rho_{\rm e}}+\frac{\hat{v'_x}}{{L_{\rho}}_x}+\frac{\hat{v'_z}}{{L_{\rho}}_z}+i\tilde{K}_x\hat{v'_x}+iK_y\hat{v'_y}+i\tilde{K}_z\hat{v'_z}=0,
\label{eq1.13}
\end{equation}
\begin{equation}
\partial_{t}\hat{v'_x}= 2\Omega \hat{v'_y}-i\tilde{K}_x \frac{\hat{P'}}{{\rho}_{\rm e}}+\frac{c_s^2}{{L_P}_x}\frac{\hat{\rho'}}{{\rho}_{\rm e}},
\label{eq1.14}
\end{equation}
\begin{equation}
\partial_{t}\hat{v'_y}=-(2-\tilde{q})\Omega \hat{v'_x}-\bar{A}_z\hat{v'_z}-i{K_y}\frac{\hat{P'}}{{\rho}_{\rm e}},
\label{eq1.15}
\end{equation}
\begin{equation}
\partial_{t}\hat{v'_z}=-i\tilde{K}_z\frac{\hat{P'}}{{\rho}_{\rm e}}-\frac{\hat{\rho'}}{{\rho}_{\rm e}}g_z,
\label{eq1.16}
\end{equation}
\begin{equation}
\partial_{t}\frac{\hat{P'}}{\rho_{\rm e}}-c_s^2\partial_{t}\frac{\hat{\rho'}}{\rho_{\rm e}}+c_s^2\frac{\hat{v'_x}}{{L_S}_x}+c_s^2\frac{\hat{v'_z}}{{L_S}_z}=0.
\label{eq1.16a}
\end{equation}
The radial and vertical length scales for pressure, density and entropy are defined by
\begin{equation}
\frac{1}{{L_P}_x}\equiv\frac{\partial_{x}P_{\rm e}}{\gamma P_{\rm e}}=\frac{1}{{L_{\rho}}_x}+\frac{1}{{L_S}_x}\equiv\frac{\partial_{x}{\rho}_{\rm e}}{ {\rho}_{\rm e}}+\frac{\partial_{x}S_{\rm e}}{\gamma S_{\rm e}},
\label{eq1.18}
\end{equation}
\begin{equation}
\frac{1}{{L_P}_z}\equiv\frac{\partial_{z}P_{\rm e}}{\gamma P_{\rm e}}=\frac{1}{{L_{\rho}}_z}+\frac{1}{{L_S}_z}\equiv\frac{\partial_{z}{\rho}_{\rm e}}{ {\rho}_{\rm e}}+\frac{\partial_{z}S_{\rm e}}{\gamma S_{\rm e}}.
\label{eq1.19}
\end{equation}
In the Boussinesq approximation equations~(\ref{eq1.13}) and~(\ref{eq1.16a}) become
\begin{equation}
\tilde{K}_x\hat{v'_x}+K_y\hat{v'_y}+\tilde{K}_z \hat{v'_z}=0,
\label{eq1.13a}
\end{equation}
\begin{equation}
\partial_{t}\frac{\hat{\rho'}}{\rho_{\rm e}}=\frac{\hat{v'_x}}{{L_S}_x}+\frac{\hat{v'_z}}{{L_S}_z},
\label{eq1.16b}
\end{equation}
Equations~(\ref{eq1.14})-(\ref{eq1.16}),~(\ref{eq1.13a}) and~(\ref{eq1.16b}) are equations (59)-(63) of Knobloch \& Spruit (1986) expressed in terms of the short wavelength shearing modes~(\ref{eq1.11}). The only difference is given by the presence of the radial stratification term in~(\ref{eq1.14}). We will see in the following, however, that it has essentially no influence on the perturbations evolution.\\
Now by deriving with respect to time the incompressibility condition~(\ref{eq1.13a}) and then, in the equation obtained, expressing $\partial_{t}\hat{v'_x}$, $\partial_{t}\hat{v'_y}$ and $\partial_{t}\hat{v'_z}$ with equations~(\ref{eq1.14}),~(\ref{eq1.15}) and~(\ref{eq1.16}),
we can express $\hat{P'}$ in terms of $\hat{\rho'}$, $\hat{v'_x}$, $\hat{v'_y}$ and $\hat{v'_z}$
\begin{eqnarray}
i\frac{\hat{P'}}{{\rho}_{\rm e}}=\frac{1}{\tilde{K}^2}[(\tilde{K}_x\frac{c_s^2}{{L_P}_x}-g_z \tilde{K}_z)\frac{\hat{\rho'}}{\rho_{\rm e}}+ \nonumber \\ 2(\tilde{q}-1)\Omega k_y \hat{v'_x} + 2\Omega \tilde{K}_x \hat{v'_y}-2 \bar{A}_z K_y \hat{v'_z}],
\label{eq1.20}
\end{eqnarray}
where ${\tilde{K}^2}={\tilde{K}_x^2}+K_y^2+\tilde{K}_z^2$. By means of equation~(\ref{eq1.20}) we obtain the system
\begin{eqnarray}
\partial_{t}\hat{v'_x}= - 2(\tilde{q}-1)\Omega\frac{ K_y \tilde{K}_x }{{\tilde{K}^2}}\hat{v'_x} + 2\Omega  (1-\frac{{\tilde{K}_x^2}}{{\tilde{K}^2}})\hat{v'_y} + \nonumber \\ 2\bar{A}_z\frac{ \tilde{K}_x  K_y}{{\tilde{K}^2}}\hat{v'_z}+\frac{c_s^2}{{L_P}_x}(1-\frac{{\tilde{K}_x^2}}{{\tilde{K}^2}}) \frac{\hat{\rho'}}{{\rho}_{\rm e}}+g_z\frac{ \tilde{K}_z \tilde{K}_x }{{\tilde{K}^2}} \frac{\hat{\rho'}}{{\rho}_{\rm e}},
\label{eq1.21}
\end{eqnarray}
\begin{eqnarray}
\partial_{t}\hat{v'_y}=\Omega [\tilde{q}-2-2(\tilde{q}-1)\frac{K_y^2}{{\tilde{K}^2}}]\hat{v'_x}- 2\Omega\frac{ K_y \tilde{K}_x}{{\tilde{K}^2}}\hat{v'_y}+ \nonumber \\ 2\bar{A}_z\frac{K_y^2}{{\tilde{K}^2}}\hat{v'_z}-\bar{A}_z\hat{v'_z}-\frac{K_y}{{\tilde{K}^2}}(\tilde{K}_x\frac{c_s^2}{{L_P}_x}-g_z \tilde{K}_z)\frac{\hat{\rho'}}{{\rho}_{\rm e}},
\label{eq1.22}
\end{eqnarray}
\begin{eqnarray}
\partial_{t}\hat{v'_z}= - 2(\tilde{q}-1)\Omega\frac{ K_y \tilde{K}_z }{{\tilde{K}^2}}\hat{v'_x} - 2\Omega  \frac{{\tilde{K}_x\tilde{K}_z}}{{\tilde{K}^2}}\hat{v'_y} + \nonumber \\2\bar{A}_z\frac{ K_y \tilde{K}_z }{{\tilde{K}^2}}\hat{v'_z}-g_z(1-\frac{{\tilde{K}_z^2}}{{\tilde{K}^2}}) \frac{\hat{\rho'}}{{\rho}_{\rm e}}-\frac{c_s^2}{{L_P}_x}\frac{ \tilde{K}_z \tilde{K}_x }{{\tilde{K}^2}} \frac{\hat{\rho'}}{{\rho}_{\rm e}},
\label{eq1.23}
\end{eqnarray}
\begin{equation}
\partial_{t}\frac{\hat{\rho'}}{\rho_{\rm e}}=\frac{\hat{v'_x}}{{L_S}_x}+\frac{\hat{v'_z}}{{L_S}_z}.
\label{eq1.24}
\end{equation}
Normalizing time with ${\Omega}^{-1}$, velocities with ${L_S}_z \Omega$ and density with $\rho_{\rm e}$  we obtain for the evolution of the non-dimensional variables $v_x$, $v_y$, $v_z$, $\rho$ the system
\begin{eqnarray}
\partial_{t}{v_x}= - 2(\tilde{q}-1)\frac{ k_y \tilde{k}_x }{{\tilde{k}^2}}{v_x} + 2 (1-\frac{{\tilde{k}_x^2}}{{\tilde{k}^2}}){v_y} +\nonumber \\  2{A_z}\frac{  \tilde{k}_x k_y }{{\tilde{k}^2}}{v_z}+ \frac{{L_S}_x}{{L_S}_z}Ri_x(\frac{{\tilde{k}_x^2}}{{\tilde{k}^2}}-1) {{\rho}}+Ri_z\frac{ \tilde{k}_z \tilde{k}_x }{{\tilde{k}^2}} {{\rho}},
\label{eq1.25}
\end{eqnarray}
\begin{eqnarray}
\partial_{t}{v_y}= [\tilde{q}-2-2(\tilde{q}-1)\frac{k_y^2}{{\tilde{k}^2}}]{v_x}- 2\frac{ k_y \tilde{k}_x}{{\tilde{k}^2}}{v_y}+\nonumber \\  2{A_z}\frac{k_y^2}{{\tilde{k}^2}}{v_z}-{A_z}{v_z}+ \frac{k_y}{{\tilde{k}^2}}(\tilde{k}_x\frac{{L_S}_x}{{L_S}_z}Ri_x+Ri_z \tilde{k}_z){{\rho}},
\label{eq1.26}
\end{eqnarray}
\begin{eqnarray}
\partial_{t}{v_z}= - 2(\tilde{q}-1)\frac{ k_y \tilde{k}_z }{{\tilde{k}^2}}{v_x} - 2  \frac{{\tilde{k}_x\tilde{k}_z}}{{\tilde{k}^2}}{v_y} + \nonumber \\  2{A_z}\frac{ k_y \tilde{k}_z }{{\tilde{k}^2}}{v_z}+ Ri_z(\frac{{\tilde{k}_z^2}}{{\tilde{k}^2}}-1) {{\rho}}+\frac{{L_S}_x}{{L_S}_z}Ri_x\frac{ \tilde{k}_z \tilde{k}_x }{{\tilde{k}^2}} {{\rho}},
\label{eq1.27}
\end{eqnarray}
\begin{eqnarray}
\partial_{t}{{\rho}}=\frac{{L_S}_z}{{L_S}_x}v_x+{v_z},
\label{eq1.28}
\end{eqnarray}
where $(k_x,{k}_y,k_z)\equiv{{L_S}_z}(K_x,{K}_y,K_z)$, $\tilde{k}_x\equiv{{L_S}_z}\tilde{K}_x$, $\tilde{k}_z\equiv{{L_S}_z}\tilde{K}_z$ and ${{\tilde{k}^2}}\equiv{{L_{S_z}^2}}{{\tilde{K}^2}}$. We introduced, as well, $A_z\equiv\frac{\bar{A}_z}{\Omega}$ and the Richardson numbers
\begin{equation}
Ri_x\equiv\frac{N_x^2}{{\Omega}^2}, \hspace{2cm} Ri_z\equiv\frac{N_z^2}{{\Omega}^2}.
\label{eq1.29}
\end{equation}
$N_x$ and $N_z$ are the Brunt-V\"{a}is\"{a}l\"{a} frequencies
\begin{equation}
N_x^2\equiv-\frac{c_s^2}{{L_S}_x{L_P}_x},  \hspace{0.5cm} N_z^2\equiv\frac{g_z}{{L_S}_z}=-\frac{c_s^2}{{L_S}_z{L_P}_z}.
\label{eq1.30}
\end{equation}
We notice that the above equations are scale invariant in the sense that results pertaining to wavenumbers $k_x$, $k_y$ and $k_z$ hold as well for wavenumbers $\beta k_x$, $\beta k_y$ and $\beta k_z$, where $\beta$ is a real number. This symmetry is broken if we introduce the effect of thermal diffusivity in the system above.\\
\section{Axisymmetry}
For axisymmetric perturbations (i.e. $k_y=0$) equations~(\ref{eq1.25})-(\ref{eq1.28}) become
\begin{equation}
\partial_{t}{v_x}=  2 \frac{{{k}_z^2}}{{{k}^2}}{v_y} - \frac{{L_S}_x}{{L_S}_z}Ri_x\frac{{{k}_z^2}}{{{k}^2}}{{\rho}}+Ri_z\frac{ k_z {k}_x }{{{k}^2}} {{\rho}},
\label{eq1.31}
\end{equation}
\begin{equation}
\partial_{t}{v_y}= (\tilde{q}-2){v_x}-{A_z}{v_z},
\label{eq1.32}
\end{equation}
\begin{equation}
\partial_{t}{v_z}= - 2  \frac{{{k}_x{k}_z}}{{{k}^2}}{v_y} -Ri_z\frac{{{k}_x^2}}{{{k}^2}} {{\rho}}+\frac{{L_S}_x}{{L_S}_z}Ri_x\frac{ {k}_z {k}_x }{{{k}^2}} {{\rho}},
\label{eq1.33}
\end{equation}
\begin{equation}
\partial_{t}{{\rho}}=\frac{{L_S}_z}{{L_S}_x}v_x+{v_z}.
\label{eq1.34}
\end{equation}
In the above equations there are no time dependent coefficients and therefore we can assume an exponential form of the type $e^{st}$ for the velocity and density perturbation fields. Neglecting radial stratification the dispersion relation reads
\begin{equation}
s^2+\frac{2(2-\tilde{q})k_z^2}{k^2}+\frac{Ri_z k_x^2}{k^2}-\frac{2k_x k_z A_z}{k^2}=0.
\label{eq1.35}
\end{equation}
This is the shearing sheet equivalent of equation (9) in Urpin (2003) in the case of zero diffusivity. For finite $Ri_z$ the vertical shear term is dominated either by the epycyclic frequency term or by the $Ri_z$ term.
In the case of weak vertical stratification (i.e. $Ri_z\ll1$) and for $\tilde{q}=3/2$ the maximum growth rate of the instability occurs for
\begin{equation}
\frac{k_x}{k_z}=\frac{\frac{1}{A_z}\pm\sqrt{\frac{1}{A_z^2}+4}}{2},
\label{eq1.36}
\end{equation}
which is the same as equation (34) of Urpin (2003) since $\frac{1}{A_z^2}\gg4$ and $\bar{A}_z=r\partial_{z}\Omega$.
The growth rate is given by
\begin{equation}
s_{\rm max}=\sqrt{\frac{2\sqrt{{1}+4{A_z^2}}}{4+\frac{1}{A_z^2}+\frac{1}{A_z}\sqrt{\frac{1}{A_z^2}+4}}} .
\label{eq1.37}
\end{equation}
For $A_z^2\ll 1$, a condition to be expected in astrophysical discs, we have therefore $s_{\rm max}\approx \vert A_z \vert$.\\
Next we consider the opposite limit by neglecting the vertical stratification. In this case the dispersion relation becomes
\begin{equation}
s^2+\frac{k_z^2}{k^2}[2(2-\tilde{q})+Ri_x]-\frac{2k_x k_z}{k^2}{A_z}=0.
\label{eq1.35A}
\end{equation}
Being radial stratification typically weak, stability is not influenced.\\
When both radial and vertical gradients are present the dispersion relation reads
\begin{eqnarray}
s^2+\frac{k_z^2}{k^2}[2(2-\tilde{q})+Ri_x]-\frac{k_x k_z}{k^2}(2{A_z}+ \nonumber\\ \frac{{L_S}_x}{{L_S}_z}Ri_x+\frac{{L_S}_z}{{L_S}_x}Ri_z)+\frac{Ri_z k_x^2}{k^2}=0.
\label{eq1.35B}
\end{eqnarray}
In this case additional destabilization is possible in principle due to the term $\Sigma=\frac{{L_S}_x}{{L_S}_z}Ri_x+\frac{{L_S}_z}{{L_S}_x}Ri_z$. However its effect will be dominated either by the stabilizing effect of the epicyclic frequency (for $k_z\geqslant k_x$) or by the last term on the LHS of~(\ref{eq1.35B}) (for $k_x\gg k_z$).
We notice that for a barotropic equilibrium ($A_z=0$) we have $\Sigma=\pm2\sqrt{Ri_x Ri_z}$ and $Ri_x$, $Ri_z$ are bound to have the same sign (Volponi 2010) and
\begin{equation}
s^2=-\frac{{k_z^2}2(2-\tilde{q})+(k_z\sqrt{Ri_x }\pm k_x\sqrt{Ri_z})^2}{k^2}
\end{equation}
In the presence of diffusion the only modification of the evolution equations occurs in~(\ref{eq1.16b}), which becomes
\begin{equation}
\partial_{t}\frac{\hat{\rho'}}{\rho_{\rm e}}=\frac{\hat{v'_x}}{{L_S}_x}+\frac{\hat{v'_z}}{{L_S}_z}-\chi {\tilde{K}^2}\frac{\hat{\rho'}}{\rho_{\rm e}},
\label{eq1.43}
\end{equation}
where $\chi$ is the thermal diffusion coefficient.\\
Once  normalized the above equation reads
\begin{equation}
\partial_{t}{{\rho}}=\frac{{L_S}_z}{{L_S}_x}v_x+{v_z}-{\tilde{k}^2}\frac{1}{Pe}{{\rho}},
\label{eq1.44}
\end{equation}
where $Pe=\frac{{{L_S^2}_z}\Omega}{\chi}$ is the Peclet number.\\
The axisymmetric dispersion relation then  becomes
\begin{eqnarray}
s^3+k^2 Pe^{-1}s^2+s\bigg[\frac{k_z^2}{k^2}[2(2-\tilde{q})+Ri_x]- \nonumber \\  \frac{k_x k_z}{k^2}(2{A_z}+\frac{{L_S}_x}{{L_S}_z}Ri_x+  \frac{{L_S}_z}{{L_S}_x}Ri_z)+\frac{Ri_z k_x^2}{k^2}\bigg]+  \nonumber \\  Pe^{-1}[2k_z^2(2-\tilde{q})-{2k_x k_z}{A_z}]=0.
\label{eq1.35C}
\end{eqnarray}
The above equation is completely equivalent to the dispersion relation (7) in Urpin (2003)  when neglecting the effect of viscosity and the same instability conditions derived in his paper follow. We briefly sum up his results. By casting equation~(\ref{eq1.35C}) in the form
\begin{equation}
s^3+a_2s^2+a_1s+a_0=0,
\label{eq1.35D}
\end{equation}
where
\begin{eqnarray}
a_2=k^2 Pe^{-1}, \nonumber \\ a_1=\bigg[\frac{k_z^2}{k^2}[2(2-\tilde{q})+Ri_x]-\frac{k_x k_z}{k^2}(2{A_z}+\nonumber \\  \frac{{L_S}_x}{{L_S}_z}Ri_x+\frac{{L_S}_z}{{L_S}_x}Ri_z)+\frac{Ri_z k_x^2}{k^2}\bigg], \nonumber\\
a_0=Pe^{-1}[2k_z^2(2-\tilde{q})-{2k_x k_z}{A_z}],
\label{eq1.35E}
\end{eqnarray}
instability conditions read (see Urpin 2003 and references therein)
\begin{equation}
a_0<0, \hspace{1cm} a_1 a_2<a_0, \hspace{1cm} a_2<0.
\label{eq1.35F}
\end{equation}
The first of the inequalities above (equation (17) in Urpin 2003),
\begin{equation}
Pe^{-1}[2k_z^2(2-\tilde{q})-{2k_x k_z}{A_z}]<0,
\end{equation}
shows that the presence of thermal diffusion relaxes the instability condition with respect to the ideal case. \\
The second inequality (equation (14) in Urpin 2003) reads
\begin{equation}
\bigg[\frac{k_z^2}{k^2}Ri_x-\frac{k_x k_z}{k^2}(\frac{{L_S}_x}{{L_S}_z}Ri_x+\frac{{L_S}_z}{{L_S}_x}Ri_z)+\frac{Ri_z k_x^2}{k^2}\bigg]<0
\end{equation}
and it is a statement about the convective stability of the disk (vertical shear does not appear in it). 

\begin{figure}
\epsfxsize=3.0in
\epsfysize=2.0in
	\begin{center}
		\includegraphics[scale=1]{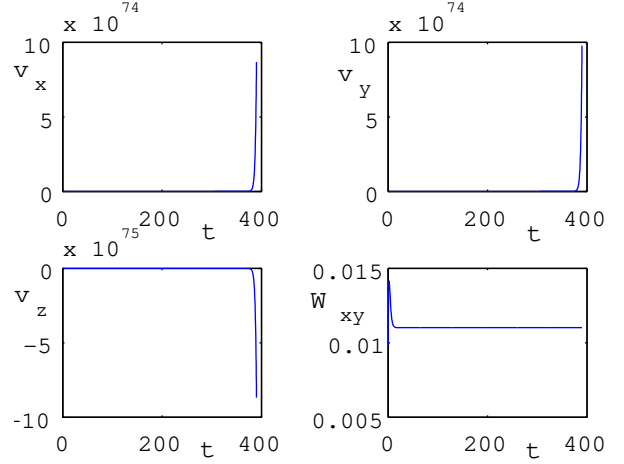}

	\end{center}
	\caption{Evolution of velocities and normalized $xy$-Reynolds stress (i.e. $W_{xy}\equiv(v_x v_y)/ v^2$, where $v^2=v_x^2+v_y^2+v_z^2$) for $Pe=10^7$, $Ri_z=-0.2$, $Ri_x=0.01$, ${L_S}_z/{L_S}_x=0.223$, $A_z=0.1$ and $k_x=500$, $k_z=50$, $k_y=0$.}
\label{axifig:1}
\end{figure}

\begin{figure}
\epsfxsize=3.0in
\epsfysize=2.0in
	\begin{center}
		\includegraphics[scale=1]{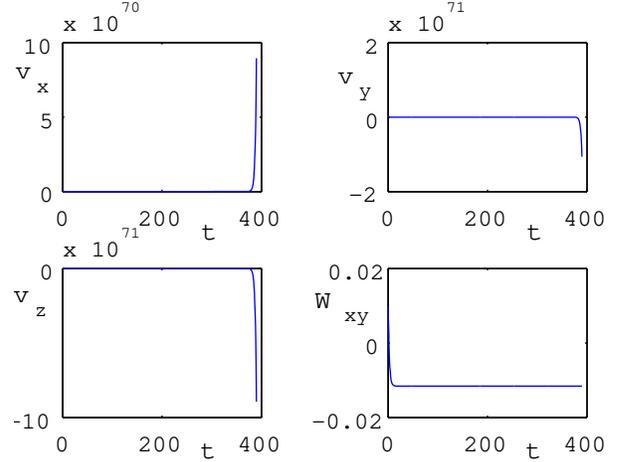}

	\end{center}
	\caption{Same as previous figure but with $A_z=0$.}
\label{axifig:2}
\end{figure}

To end this section we present the evolution of axisymmetric perturbations for a convectively unstable vertical stratification. In Fig.~\ref{axifig:1} we set  $k_x/k_z=10$, which pertains to the maximal vertical shear growth rate for $A_z=0.1$, $Ri_z=-0.2$ and $Pe=10^7$. 
We notice that the growth rate is the convective one ($\sqrt{-Ri_z}$), but differently from conventional understanding $W_{xy} \equiv (v_x v_y)/ v^2>0$, where $v^2=v_x^2+v_y^2+v_z^2$. To ascertain the origin of the positive sign of $W_{xy}$  we set $A_z=0$. As can be seen in Fig.~\ref{axifig:2}, this results in $W_{xy}<0$.\\
We can therefore conclude that, when the vertical convective growth rate is larger than the vertical shear one, the ensuing instability is of mixed type in the sense that the growth rate is given by convection whereas the sign of $W_{xy}$ is determined by vertical shear. \\
Linear inviscid axisymmetric perturbations cannot transport angular momentum in hydrodynamic discs (Ruden, Papaloizou \& Lin 1988), however we will see in the next section that for non-axisymmetric perturbations, which induce transport, the same type of behaviour described above holds. \\
We determined as well the effect of thermal diffusion on the growth rate $s$. In Table~\ref{tab:1} we report the dependence of $s$ on $Pe$. We notice that $s$ is drastically reduced down to $s\sim A_z$. Further decreasing of $Pe$ does not have effect on $s$. 
\begin{table}
\caption {Dependence of mixed (convective + vertical shear) axisymmetric growth rate on Peclet number (relative to the case in Fig. 1)}  
\label{tab:1} 
\begin{center}
    \begin{tabular}{ | l | p{0.7cm} | p{0.7cm} | p{0.7cm} | p{0.7cm} | p{0.7cm} | p{0.7cm} |}
    \hline
    $Pe$ &  $\infty$ &  $10^7$ & $10^6$  & $10^5$ & $10^4$ & $10^3$\\ \hline
    $s$ & $0.44$ & $0.44$ & $0.35$ & $0.14$ & $0.1$ & $0.1$\\ \hline
    \end{tabular}
\end{center}
\end{table}
\section{Non-axisymmetry}
\subsection{Case 1: only vertical shear drive}
\begin{figure}
\epsfxsize=3.0in
\epsfysize=2.0in
	\begin{center}
		\includegraphics[scale=1]{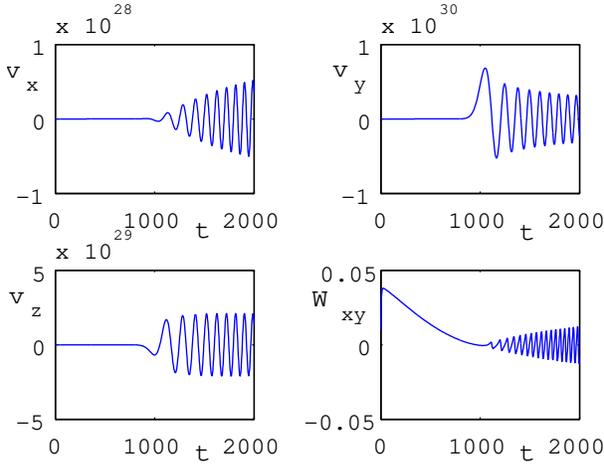}

	\end{center}
	\caption{Evolution of velocities and normalized $xy$-Reynolds stress in the case of very weak stratification for $A_z=0.1$ and $k_x=1000$, $k_z=100$, $k_y=1$.}
\label{fig:1}
\end{figure}

We consider here the evolution of ideal non-axisymmetric perturbations in the presence of a convectively stable vertical stratification.\\ 
With a reference stratification of $Ri_z=0.1$, $Ri_x=0.01$ and a realistic $A_z$, perturbations always decay for any value of the wavenumbers. In this case the vertical shear destabilizing drive is never able to overcome the stabilizing influence of stratification and epicyclic frequency. \\
If we allow for weaker stratification the vertical shear can drive huge transient growths in the perturbations.\\
Fixing $A_z$ to a positive value, the interval of growth is connected to $\tilde{k}_x$ and $\tilde{k}_z$ having the same sign. We can see this in Fig.~\ref{fig:1} for the case $\tilde{k}_x>0$ and $\tilde{k}_z>0$, where we set the $k_x/k_z$ ratio to the value pertaining to the maximal axisymmetric growth rate $A_z$. By increasing $k_y$ clearly the interval of growth is shortened  and therefore the largest amplification reached diminishes. We stress that the transients are indistinguishable from a full-fledged instability. The growths are in fact exponential with growth rate close to $A_z$. Varying the ratio $k_x/k_z$, as well, causes a decrease in the maximum amplification reached. \\
In the case $A_z<0$  growths occurs when $\tilde{k}_x$ and $\tilde{k}_z$ have opposite signs. \\
We notice as well that during the phase of growth, radial transport is always positive. \\
As stated before eventually all perturbations decay. Heuristically this can be shown neglecting the stratification in equations~(\ref{eq1.25})-(\ref{eq1.28}). By using the incompressibility condition~(\ref{eq1.13a}) we obtain for the $x$ and $y$ velocity components the evolution equations
\begin{eqnarray}
\partial_{t}{v_x}= - 2(\tilde{q}-1+{A_z}\frac{\tilde{k}_x }{{\tilde{k}_z}})\frac{ k_y \tilde{k}_x }{{\tilde{k}^2}}{v_x} + \nonumber \\ 2 (1-\frac{{\tilde{k}_x^2}}{{\tilde{k}^2}}-{A_z}\frac{ k_y^2 \tilde{k}_x }{{\tilde{k}^2}\tilde{k}_z}){v_y} ,
\label{eq1.38}
\end{eqnarray}
\begin{eqnarray}
\partial_{t}{v_y}= [\tilde{q}-2-2(\tilde{q}-1)\frac{k_y^2}{{\tilde{k}^2}}-2{A_z}\frac{k_y^2 \tilde{k}_x }{{ \tilde{k}^2 \tilde{k}_z }}+\nonumber \\ {A_z}\frac{ \tilde{k}_x }{{ \tilde{k}_z }}]{v_x}+({A_z}\frac{k_y}{{ \tilde{k}_z}}- 2\frac{ k_y \tilde{k}_x}{{\tilde{k}^2}}-2{A_z}\frac{k_y^3}{{\tilde{k}^2 \tilde{k}_z}}){v_y},
\label{eq1.39}
\end{eqnarray}
which in the limit $t\longrightarrow\infty$ become
\begin{equation}
\partial_{t}{v_x}=  2\frac{A_z^2}{q^2+A_z^2}{v_y} ,
\label{eq1.40}
\end{equation}
\begin{equation}
\partial_{t}{v_y}= -2{v_x}.
\label{eq1.41}
\end{equation}
These can be reduced to the equation
\begin{equation}
\partial_{t}^2{v_x}=  -4\frac{A_z^2}{q^2+A_z^2}{v_x} ,
\label{eq1.42}
\end{equation}
which implies stability.

For sake of completeness, we considered as well the case of stable weak vertical and unstable strong radial stratifications. The vertical shear growth rate is only slightly enhanced.
\subsection{Case 2: combined effect of vertical shear and vertical unstable convection drives}
\begin{figure}
\epsfxsize=3.0in
\epsfysize=2.0in
	\begin{center}
		\includegraphics[scale=1]{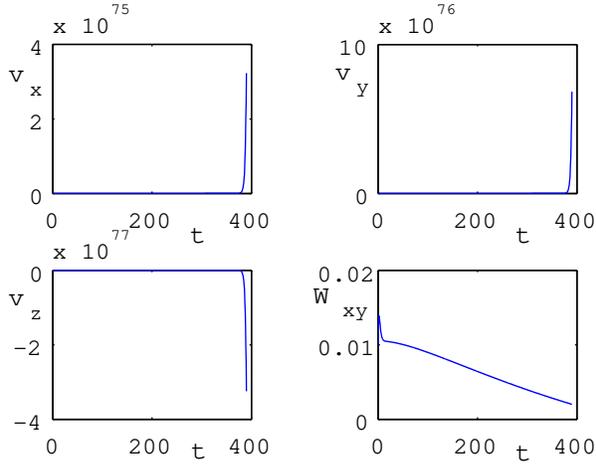}

	\end{center}
	\caption{Evolution of velocities and normalized $xy$-Reynolds stress (up to $t=400$)  for $Ri_z=-0.2$, $Ri_x=0.01$, ${L_S}_z/{L_S}_x=0.223$, $A_z=0.1$ and $k_x=500$, $k_z=50$, $k_y=1$.}
\label{fig:2}
\end{figure}

\begin{figure}
\epsfxsize=3.0in
\epsfysize=2.0in
	\begin{center}
		\includegraphics[scale=1]{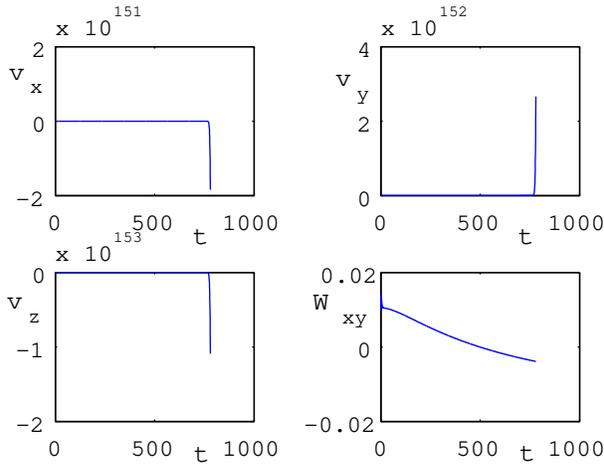}
	\end{center}
	\caption{Same as previous picture up to $t=800$.}
\label{fig:3}
\end{figure}

\begin{figure}
\epsfxsize=3.0in
\epsfysize=2.0in
	\begin{center}
		\includegraphics[scale=1]{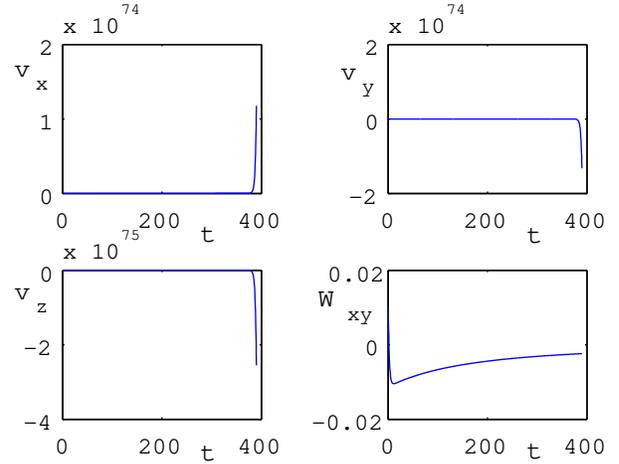}
	\end{center}
	\caption{Evolution of velocities and normalized $xy$-Reynolds stress (up to $t=400$) for $Ri_z=-0.2$, $Ri_x=0.01$, ${L_S}_z/{L_S}_x=0.223$, $A_z=0$ and $k_x=500$, $k_z=50$, $k_y=1$.}
\label{fig:4}
\end{figure}

We examine here the evolution of ideal perturbations in the presence of an unstable vertical stratification. This is the central part of this study, because the interaction of vertical shear and convective drives results in evolutions which are determined by the relative strength of the vertical shear growth rate ($s_{\rm vs} \sim \vert A_z \vert$) and the vertical convective growth rate ($s_{\rm c}\sim\sqrt{-Ri_z}$). For $s_{\rm vs}>s_{\rm c}$ we observed the same type of behaviour described in Case 1 - large exponential amplifications with growth rate $s_{\rm vs}$  occur with asymptotic stabilization. For $s_{\rm c}>s_{\rm vs}$ the system is asymptotically unstable with growth rate $s_{\rm c}$. For $s_{\rm vs} \sim s_{\rm c}$ a variety of mixed evolutions is possible: these include transient growths, exponential instability and algebraic instability.\\
In the following we will concentrate on the case $s_{\rm c}>s_{\rm vs}$, for which the system is asymptotically unstable. Similarly to what we found in the axisymmetric case, the main difference with a purely convective instability is given by the sign of the angular momentum transport, which is positive exactly in the same time intervals in which transient growth occured in Case 1. In other words for $A_z>0(<0)$ positive transport occurs when $\tilde{k}_x\tilde{k}_z>0(<0)$.\\
This is shown in Figs.~\ref{fig:2} and~\ref{fig:3} where we fixed $k_x/k_z\sim10$ pertaining to maximal axisymmetric vertical shear growth rate for $A_z=0.1$ ($s_{\rm vs}\sim 0.1$), and chose $Ri_z=-0.2$. As can be seen, the growth rate is given by $s_{\rm c}\sim \sqrt{-Ri_z}$ and transport is positive in the regimes described above. This is important because, in the absence of vertical shear, convection has the tendency to transport angular momentum inward rather than outward (Ryu \& Goodman 1992). To see this we switched off the vertical drive (i.e. $A_z=0$) and found that this results in negative transport (Fig.~\ref{fig:4}).\\
We stress that this mechanism occurs for any value of the vertical shear growth rate, however small.\\
We noticed as well that when $s_{\rm c}>s_{\rm vs}$, but their values are close, the growth rate of the mixed instability, $s_{\rm mix}$, is enhanced of a fraction of $s_{\rm c}$. For example, in the case $k_x/k_z\sim3$, $A_z=0.4$ (i.e. $s_{\rm vs} \sim 0.4$), $Ri_z=-0.2$ (i.e. $s_{\rm c} \sim 0.45$) and $k_y=0.1$  we obtained $s_{\rm mix} \sim 0.55$.
\subsection{Case 3: presence of thermal diffusion}
\begin{figure}
\epsfxsize=3.0in
\epsfysize=2.0in
	\begin{center}
		\includegraphics[scale=1]{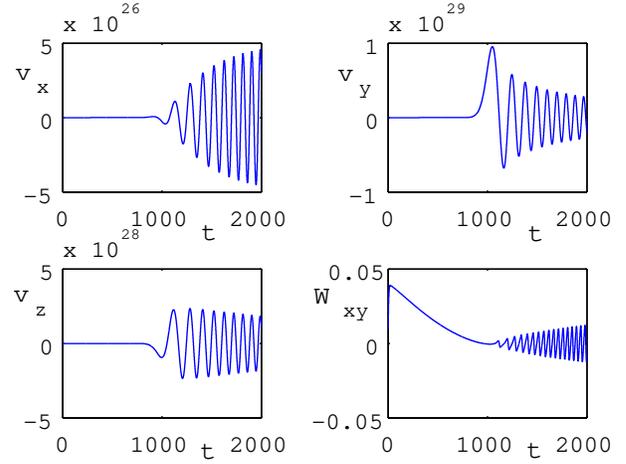}
	\end{center}
	\caption{Evolution of velocities and normalized $xy$-Reynolds stress for $Pe=10^5$, $Ri_z=0.1$, $Ri_x=0.01$, ${L_S}_z/{L_S}_x=0.316$, $A_z=0.1$ and $k_x=1000$, $k_z=100$, $k_y=1$.}
\label{fig:5}
\end{figure}

\begin{figure}
\epsfxsize=3.0in
\epsfysize=2.0in
	\begin{center}
		\includegraphics[scale=1]{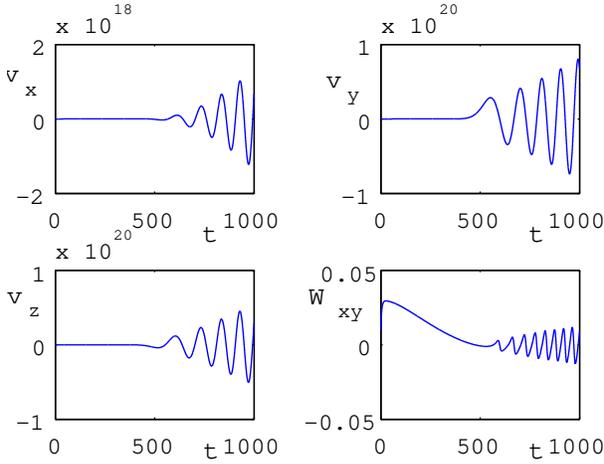}
	\end{center}
	\caption{Evolution of velocities and normalized $xy$-Reynolds stress for $Pe=10^5$, $Ri_z=-0.2$, $Ri_x=0.01$, ${L_S}_z/{L_S}_x=0.223$, $A_z=0.1$ and $k_x=500$, $k_z=50$, $k_y=1$.}
\label{fig:6}
\end{figure}

\begin{figure}
\epsfxsize=3.0in
\epsfysize=2.0in
	\begin{center}
		\includegraphics[scale=1]{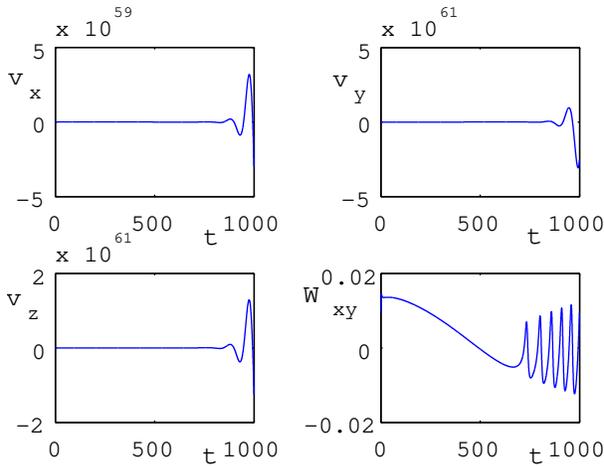}
	\end{center}
	\caption{Evolution of velocities and normalized $xy$-Reynolds stress for $Pe=10^6$, $Ri_z=-0.2$, $Ri_x=0.01$, ${L_S}_z/{L_S}_x=0.223$, $A_z=0.1$ and $k_x=500$, $k_z=50$, $k_y=1$.}
\label{fig:7}
\end{figure}

\begin{figure}
\epsfxsize=3.0in
\epsfysize=2.0in
	\begin{center}
		\includegraphics[scale=1]{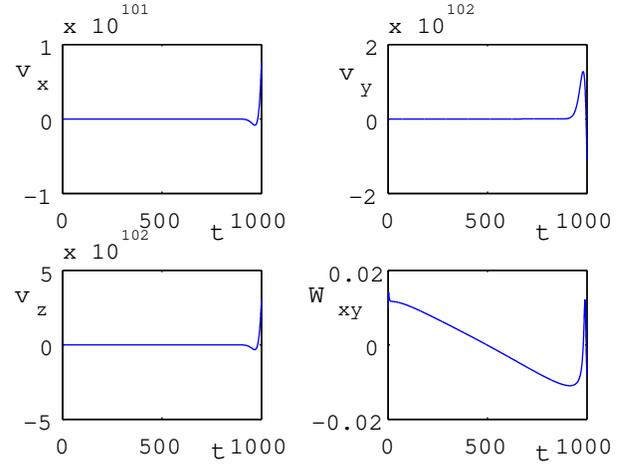}
	\end{center}
	\caption{Evolution of velocity and normalized $xy$-Reynolds stress for $Pe=10^5$, $Ri_z=-0.2$, $Ri_x=0.01$, ${L_S}_z/{L_S}_x=0.223$, $A_z=0.1$ and $k_x=100$, $k_z=10$, $k_y=0.2$.}
\label{fig:8}
\end{figure}

Two are the main differences with respect to the ideal case. \\
The first regards the vertical shear instability and consists in the fact, predicted by the axisymmetric theory (Urpin 2003), that the condition for instability is relaxed. Indeed for non-axisymmetric perturbations as well, we noticed that the presence of a stable vertical stratification does not have influence on growth, which occurs for any value of $Ri_z$. In this case also, instability transforms into a transient amplification with growth rate close to the axisymmetric one, as shown in Fig.~\ref{fig:5}. Again growth is connected to the signs of $\tilde{k}_x$ and $\tilde{k}_z$ exactly in the same way discussed in Case 1.\\
The second difference consists in the fact that, as for the axisymmetric case, thermal diffusion has a stabilizing effect on the convective instability. This can be seen in Fig.~\ref{fig:6}, where we considered $Ri_z=-0.2$, $A_z=0.1$, $k_x=500$, $k_y=1$, $k_z=50$, with $Pe=10^5$. Convective growth is suppressed and the growth rate in the amplification phase is the one pertaining to the vertical shear destabilization. 
We can however increase the growth rate toward the one pertaining to a convective instability in two ways. The first is by increasing the Peclet number as done in Fig.~\ref{fig:7}, where parameters have the same values as in Fig.~\ref{fig:6} but with $Pe=10^6$. 
The second is by increasing the wavelength of the perturbations as done in Fig.~\ref{fig:8} which is the same as Fig.~\ref{fig:6} except that now $k_x=100$, $k_z=10$, $k_y=0.2$. It can be seen that the growth rate is strongly enhanced. We caution however that here we are reaching a borderline regime for our short wavelength theory.\\
We can sum up the effects of thermal diffusivity as follows. Diffusivity acts as a stabilization mechanism for convection and not for the vertical shear instability. By decreasing the Peclet number the growth rate decreases until  the vertical shear instability growth rate is reached. There the diffusion induced stabilization stops.
On the other side if we keep $Pe$ fixed and decrease wavenumbers, the growth rate increases toward the convective instability value ($\sim\sqrt{-Ri_z}$). Therefore we can conclude that the stabilizing effect of thermal diffusion affects convection for short wavelengths.
\section{Summary}
We presented the evolution of non-axisymmetric perturbations in discs whose rotation profile is $x$ and $z$ dependent. In the shearing sheet limit and moving to a reference frame corotating with the background flow the presence of vertical shear induces the evolution of the vertical wavenumber $\tilde{k}_z$.  We considered short wavelength perturbations. 

Perturbations are always stabilized asymptotically; however, huge exponential growths can occur which are virtually indistinguishable from a full-fledged instability. The condition for the occurrence of such amplifications is $\tilde{k}_x\tilde{k}_z>0$ for $A_z>0$ and $\tilde{k}_x\tilde{k}_z<0$ for $A_z<0$. We summarized this in Table~\ref{tab:2}.\\
We examine more in detail this condition. Let's consider $A_z>0$, $k_x>0$ and $k_z>0$. In this case $\tilde{k}_x$ is always positive, $\tilde{k}_z$ instead evolves from positive to negative values. Therefore we observe the transient exponential phase of growth when $k_z-A_z k_y t>0$ (i.e. for $t<\frac{k_z}{A_z k_y}\equiv t_g)$. We stress that this is different from the usual transient growth paradigm (Chagelishvili et al. 2003) since here the growth is exponential and the physical origin is the vertical shear drive. Similar type of exponential transient growths were described by Balbus \& Hawley (1992) in the context of the non-axisymmetric magnetorotational instability, Korycansky (1992) in hydrodynamic convection and Volponi, Yoshida \& Tatsuno (2000) in the stabilization of plasmas kink modes. In all these studies, though, the flow was just radially sheared. \\
For $t\sim t_g$ our short wavelength approximation breaks down since $\tilde{k}_z\longrightarrow 0$; however, huge amplification values are reached at times $t$ well before $t_g$, when $\tilde{K}_z\gg H$.\\
The growth rate is close to the axisymmetric growth rate. The transport is positive in the phase of growth. As for the axisymmetric theory the most easily destabilizable modes occur in the non-adiabatic case of finite thermal diffusivity. In the ideal case the presence of a stable vertical stratification prevents the occurrence of growth. Growth is possible only for small values of the vertical Richardson number.

This first set of results extends the axisymmetric theory to non-axisymmetric perturbations, providing as well a linear local mechanism to explain the non-axisymmetric non-linear results in Nelson et al. (2012). The linear non-axisymmetric mechanism is essentially the axisymmetric one which is still active for short to intermediate times.
\begin{table}
\caption {Vertical Shear (Ideal, $Ri_z\ll1$, or Non ideal)}  
\label{tab:2} 
\begin{center}
    \begin{tabular}{ | l | p{2.5cm} | p{2.5cm} | }
    \hline
     &  $A_z>0$ & $A_z<0$  \\ \hline
    $\tilde{k}_x\tilde{k}_z>0$ & Transient \hspace{-0.3cm}
Exp. Growth ($W_{xy}>0$) & Decay  \\ \hline
    $\tilde{k}_x\tilde{k}_z<0$ & Decay & Transient Exp. Growth ($W_{xy}>0$)\\ \hline
    \end{tabular}
\end{center}
\end{table}
\begin{table}
\caption {Vertical Shear and Convection (Ideal, $s_{\rm c}>s_{\rm vs}$)}  
\label{tab:3} 
\begin{center}
    \begin{tabular}{ | l | p{1.5cm} | p{1.5cm} | p{1.5cm} | }
    \hline
     &  $A_z>0$ & $A_z<0$  & $A_z=0$\\ \hline
    $\tilde{k}_x\tilde{k}_z>0$ & Instability ($W_{xy}>0$)& Instability ($W_{xy}<0$)& Instability ($W_{xy}<0$)\\ \hline
    $\tilde{k}_x\tilde{k}_z<0$ & Instability ($W_{xy}<0$) & Instability ($W_{xy}>0$) & Instability ($W_{xy}<0$) \\ \hline
    \end{tabular}
\end{center}
\end{table}

We considered as well the concomitant action of the vertical shear and vertical convective instabilities. The asymptotic behaviour depends on the relative strength of $s_{\rm vs}$ and $s_{\rm c}$.\\
For $s_{\rm vs}>s_{\rm c}$ we observed the same type of behaviour described above - large growths occur with asymptotic stabilization. When $s_{\rm c}>s_{\rm vs}$ the system is asymptotically unstable. For $s_{\rm vs} \sim s_{\rm c}$ a variety of mixed evolutions is possible.\\
The most interesting feature for the present study is connected with the sign of the angular momentum transport. In fact we observed that in the phase in which the vertical shear driven transients growths occur (i.e. $\tilde{k}_x\tilde{k}_z>0$ for $A_z>0$ and $\tilde{k}_x\tilde{k}_z<0$ for $A_z<0$) transport is always positive, even in the case $s_{\rm c}>s_{\rm vs}$. When the vertical shear drive is switched off (i.e. $A_z=0$), the convective transport turns negative. Table~\ref{tab:3} summarizes these results.\\
We conclude that, for $s_{\rm c}>s_{\rm vs}$, the interaction of vertical shear growths and convective instability results in a destabilization whose strength (i.e. growth rate) is governed mainly by convection and whose transport by vertical shear. This mechanism holds for any value of $s_{\rm vs}(<s_{\rm c})$, however small. Again in the case $s_{\rm c}>s_{\rm vs}$, but when their values are close, the growth rate of the mixed instability is increased of a fraction of $s_{\rm c}$ with respect to $s_{\rm c}$.

Thermal diffusion has a stabilizing influence on the convective instability. This is specially strong for short wavelengths. Instead it has no influence on the vertical shear growth rate. In the case $s_{\rm c}>s_{\rm vs}$, by decreasing the Peclet number, we observed the transition from convection to vertical shear dominated evolution. By increasing the Peclet number or decreasing the wavenumbers huge growths occur. The regime of very long wavelengths is outside the scope of the present theory.\\
We can say that whereas non-axisymmetricity stabilizes the axisymmetric vertical shear instability asymptotically for $t\longrightarrow \infty$ (not for intermediate times though) but has weak influence on the convective one, non-adiabaticity stabilizes the convective instability but has no influence on the vertical shear growth rate.

This second set of results instead points at a new possible path to explain outward angular momentum transport in astrophysical discs: interaction of vertical shear and convective instabilities.
\section*{Acknowledgements}
The author would like to express his gratitude to Prof. Zensho Yoshida for his suggestions, advice  and continuous support and to Prof. Ryoji Matsumoto for the many discussions and encouragement. He would also like to acknowledge very stimulating conversations with Prof. Alexander Tevzadze.

\end{document}